\documentclass[review]{elsarticle}

\usepackage{lineno,hyperref}
\modulolinenumbers[5]

\journal{Journal of \LaTeX\ Templates}

\usepackage{amsmath}
\usepackage{amssymb}
\usepackage{graphicx}
\usepackage{dcolumn}
\usepackage{bm}

\newcommand{\epss}{\varepsilon_{\sigma}}
\newcommand{\epsv}{\varepsilon_{V}}

\newcommand{\eps}{\varepsilon}

\newcommand{\BEQ}{\begin{equation}}
\newcommand{\EEQ}{\end{equation}}
\newcommand{\BEN}{\begin{align}}
\newcommand{\EEGN}{\end{align}}
\newcommand{\BES}{\begin{subequations}}
\newcommand{\EES}{\end{subequations}}
\newcommand{\BEA}{\begin{eqnarray}}
\newcommand{\EEA}{\end{eqnarray}}

\newcommand{\omgp}{ \omega^{\prime}}

\newcommand{\omg}{ \omega}

\newcommand{\nn}{\nonumber}

\begin{document}

\begin{frontmatter}



\title{Information transfer through a 
 signaling module with feedback: a perturbative approach}


\author{Gerardo Aquino$^1$}
\author{Martin Zapotocky$^2$}
\address{$^1$ Department of Life Sciences, Imperial College,  SW7 2AZ, London, UK,\\ $^2$ Institute of Physiology of the Czech Academy of Sciences, Videnska 1083, 14220 Prague, Czech Republic.}
\date{\today }

\begin{abstract}
Signal transduction in biological cells is effected by signaling pathways that typically include multiple feedback loops. Here we analyze information transfer through a prototypical signaling module with biochemical feedback. The module switches stochastically between an inactive and active state; the input to the module governs the activation rate while the output (i.e., the product concentration) perturbs the inactivation rate. Using a novel perturbative approach, we compute the rate with which information about the input is gained from observation of the output. We obtain an explicit analytical result valid to first order in feedback strength and to second order in the strength of input. The total information gained during an extended time interval is found to depend on the feedback strength only through the total number of activation/inactivation events. 

\end{abstract}

\end{frontmatter}
\section{Introduction}
Accurate sensing of the environment is crucial for the survival of biological organisms. Bacteria, as well as animal chemoreceptor cells, can sense certain chemicals in their chemical environment with high precision, in some cases near the single-molecule detection limit \cite{BP,mao}. 
The effect of the extracellular stimulus on the cell is mediated by the signal transduction pathway - a complex biochemical reaction network. The pathway is based on a sequence of transduction steps, with each subsequent step being effected by the chemical product of the previous step. The first step typically consists in the activation of a receptor protein in the cell membrane by the external stimulus, which results in an ion influx or in the production of a second messenger chemical. This leads to the activation of subsequent steps within the cell. Through molecular feedback loops, the product of a given transduction step may regulate its own production, or influence earlier (upstream) steps of the pathway. As signal transduction pathways are inherently noisy \cite{ladbury}, faithful transmission of information through the pathways requires amplification and/or adaptation. This is often accomplished through positive (amplification) and negative (adaptation) feedback built into the reaction network. 

Recent years have brought the use of information-based measures to characterize the reliability of biological signal transduction \cite{tosto,cheong,bialek,tenwoldepre}. Such measures explicitly evaluate the amount of information about the stimulus that it is trasmitted through the signaling pathway. Mutual information between the stimulus and the pathway output (i.e., the product of the final transduction step) was evaluated in, e.g.\cite{cheong,bialek} , to assess the precision with which stationary stimuli of different strengths can be distinguished. In many signaling scenarios, however, it is important to faithfully transduce the temporal variations of the input stimulus, which encode biologically important information. Some recent studies have evaluated the transduction reliability for time-varying signals by computing the information transmission rate. In \cite{tenwoldepre}, prototypical signaling pathways with feedback were represented by coupled Langevin equations with additive Gaussian noise, and thefrequency-dependent  gain-to-noise ratio was computed. In the Gaussian noise approximation, this gain (together with the input power spectrum) determines the the mutual information between the time courses of the stimulus and of the output.  Some of the stochasticity within a signaling pathway, however, arises directly from the inherent stochastic dynamics of the transduction components, and cannot always betreated as additive noise; in such a case, the information calculation cannot be reduced to the evaluation of the gain-to-noise ratio.

In this work, we carry out a perurbative computation of information transfer through a simple prototypical signaling module with biochemical feedback. The module switches stochastically between two states, with switching rates governed by the stimulus (i.e., input) and by the product (i.e., output). No additive external noise is assumed. Such autoregulated stochastic modules arise within various signal transduction and gene regulation pathways (see Sec. 2).  In a previous investigation \cite{boro} negative feedback, in this module, was shown to decrease the signal-to-noise ratio (SNR) at the output, but at the same time to increase the spectral range of the response - thus yielding no obvious expectation on how feedback overall affectsinformation transmission through the module. Here, we address this question by directly quantifying the information that is gained about the external stimulus from following the module output. In order to achieve this we introduce a novel perturbative approach on a conveniently defined relative entropy for stochastic point processes \cite{Hanggi}. This information gain is well-defined for a single stimulus trajectory (i.e., it requires no averaging over stimuli as in mutual-information-based measures) \cite{Hanggi}. We obtain an explicit analytical result valid to first order in feedback strength and to second order in the strength of input. Surprisingly, the total information gained during a long time interval is found to be proportional to the total number of state-switching events, with no further dependence on the feedback strength or on the spectral distribution of the input. We compare this result to previous investigations of information transfer through some related information channels. 


\section{ A two-state signaling  module with feedback}
We consider a simple signaling module based on a single protein that switches between two conformational states. These may correspond to the open and closed state of an ion channel, or to the active/inactive states of an enzyme within a larger signaling network. 
We introduce the module by referring to the example of a calcium ion channel that is autoregulated by calcium-mediated feedback. When the channel is open, calcium ions flow from the extracellular space into the cell. This leads to a fast increase of the free calcium concentration in the immediate vicinity of the channel (the increase is localized as calcium buffering in the cytoplasm leads to the formation of a calcium microdomain \cite{parek08}). For certain types of calcium channels (such as the voltage-activated L-type channels \cite{pete99} or the cyclic-nucleotide-gated (CNG) channels \cite{brad05,reidl}), the cytoplasmic calcium can inactivate the channel when it binds to the channel/calmodulin complex. This implements an autoregulatory feedback loop that shortens the response to the external gating signal, and thus helps to faithfully transduce fast signal variations. Similar autoregulatory loops, in which the product of a particular step in the pathway downregulates its own production, arise in numerous signal transduction and gene regulation networks \cite{rose02}. 

The stochastic switching of the channel state is governed by the opening rate (assumed to depend on the extracellular stimulus gating the channel) and the closing rate. The negative autoregulation may be effected through a calcium-dependent increase of the closing rate or decrease of the opening rate. For the CNG channels that motivated us in this study, electrophysiological data indicates that the binding of calcium to the channel/calmodulin complex increases the closing rate. We consequently make only the closing rate depend on the calcium concentration.

The signaling module is shown schematically in Fig. 1.  When the channel is in the open state, ions flow into the microdomain at a fixed rate J. 
 Once inside the microdomain, the ions are cleared out through ion pumps or exchangers in the membrane, as well as by diffusion within the cytoplasm; we assume a first-order clearance kinetics with rate constant  $\lambda$ (see Fig. 1). 
The dynamics of the concentration c of ions in the cell compartment is given by the following equation: 
\BEQ
\label{raw}
\frac{dc(t)}{dt}=\frac{J}{\Delta}S(t) -\lambda c(t),
\EEQ
where $J$ is the flow  of ions entering the cell through the open channel, $\Delta$ the volume of the  cell micro-domain and the two-valued function  $S(t)=1$ or $0$ indicates the open or closed
state of the channel.
\begin{centering}
\begin{figure}[h!]
\hspace{2cm}\includegraphics[height=3.8 cm,width=6.8 cm, angle=0]{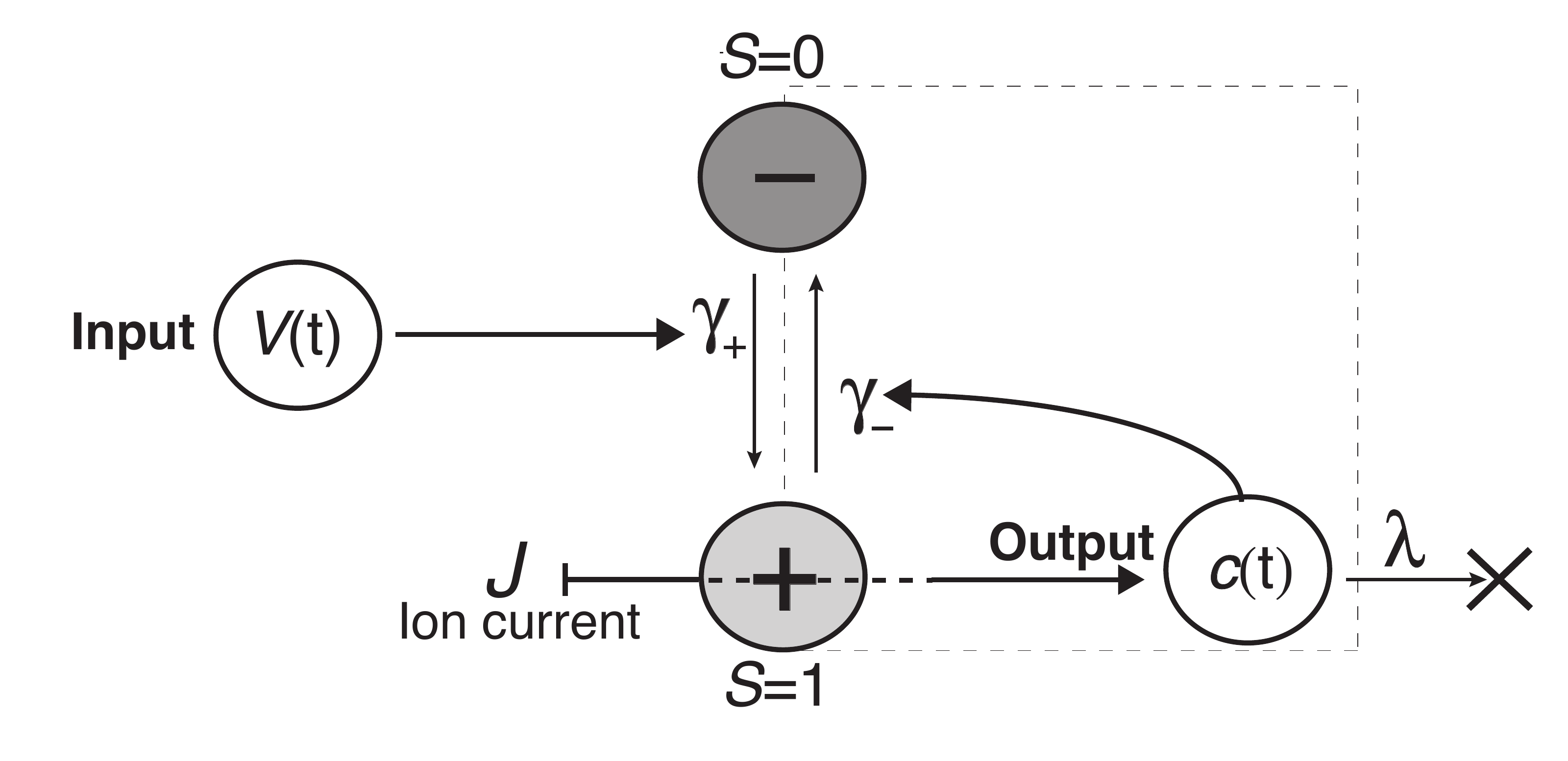}
\caption{The ion-channel (gray) opens  with  rate $\gamma_+$ governed by
the input  $ V(t)$ and closes with rate $\gamma_-$ that depends on the ion concentration $c(t)$ (which is considered to be the output).  Ions inside the micro-domain  (dashed box) are cleared out at rate $\lambda$.}
\end{figure}
\end{centering}
Adopting the dimensionless variable  $\sigma(t)=c(t) \lambda \Delta/J$, Eq.  (\ref{raw})  becomes
\BEQ
\label{dynamics}
\frac{d\sigma}{dt}=\lambda(S(t)-\sigma(t))
\EEQ
and $\sigma(t)$ is restricted to the range $[0,1]$.
Examples of  trajectories  of $\sigma(t)$  are shown in Fig. 2.
 The switching events of the channel between  time  $0$ and $t$:
\BEQ
\label{pp}
0<t_1<t_2< \dots <t_n<t
\EEQ
are the realization of a stochastic point process driving the  dynamics of $\sigma(t)$.
 In fact, fixed a sequence  $\vec{t}_n=(0,t_1,\cdots,t_n)$  of such events and the initial condition $S(0)=i$ and $\sigma(0)=\sigma_0$, an exact solution for $\sigma(t)$ follows directly from  Eq. (\ref{dynamics}): 
\begin{align}
\label{sigmaev}
\nn & \sigma(t|\vec{t}_{2n})=i+\left[\sigma_0  -i +(-1)^{i}\sum_{j=0}^{n-1}\left(e^{\lambda t_{2j}}-e^{\lambda t_{2j+1}}\right)\right]e^{-\lambda t}\\
 &  \sigma(t|\vec{t}_{2n+1})=1-i+
\left[\sigma(t=t_{2n+1}|\vec{t}_{2n})-1+i\right]e^{-\lambda (t-t_{2n+1})}. 
\end{align}
 Eqs. (\ref{sigmaev})  determine  $\sigma(t)$ after 
either an even or  an odd sequence  of switching events and show how
this variable keeps a full record (memory) of these events. 

We assume that the external input   $V(t)$  gates the channel by  perturbing only the opening rate,
i.e.
\BEQ
\tilde{\gamma}_+=\gamma_+(1 +\eps_V V(t)),
\EEQ
while  the internal ion  concentration affects the closing rate
\BEQ
\label{gammam}
\tilde{\gamma}_-=\gamma_-(1 +\eps_{\sigma} \sigma(t)).
\EEQ
When $\eps_{\sigma} >0$, an increase in calcium concentration leads to faster closing of the channel and hence a reduced calcium influx. Therefore $\eps_{\sigma} >0$  corresponds to negative autoregulation and $ \eps_{\sigma} <0$ to positive feedback on the calcium dynamics (to guarantee that 
 $\tilde{\gamma}_-$ remains positive in Eq. (\ref{gammam}), $\eps_{\sigma} >-1$ is required). Note that the opening rate in Eq. (\ref{gammam})  depends, through Eq. (\ref{sigmaev}), on the whole history of previous openings and closings of the channel. The feedback introduced through $\eps_{\sigma}$  therefore renders the channel state dynamics non-Markovian.

\section{Signal-induced information gain} 
As quantification of the information transfer   we adopt the information gained during the interval $[0,t]$  due to  the stimulus $V$  \cite{Hanggi}.
To introduce this measure of information transfer we start from the definition of entropy for a stochastic point process that generates switching events. Denoting by  $\Sigma_n(t,t_n,\dots,t_1,t_0)$ the probability density  that a sequence of $n$ events occurs at time $t_1, \dots, t_n$ between time $t_0$ and $t$ the entropy of the stochastic process  during this time interval can be written as
\begin{align}
\label{tauent}
S_{\Delta \tau}(t,t_0)&=-\Sigma(t,t_0) \ln \Sigma(t,t_0) -
\nn\sum_{n=1}^{\infty}\int_{t_0}^t dt_1\int_{t_1}^{t} dt_{2}\\ \dots& \int_{t_{n-1}}^{t}dt_n \Sigma_n(t,t_n,\dots,t_0)\ln[\Sigma_n(t,t_n,\dots,t_0)(\Delta \tau)^n]
\end{align}
Eq. (\ref{tauent})  is the natural extension of the Shannon entropy to a continuous-time process assuming a finite resolution $\Delta \tau$ with which the continuous-time variable can be measured. This $\tau$-entropy, as well as the difference of such entropies \cite{Hanggi}, depends on $\Delta \tau$ and is consequently ill-defined (not fully defined by the stochastic process). To reach a uniquely defined measure of information transfer, Goychuk and Hanggi \cite{Hanggi}, introduced the relative entropy (or information gain) $\mathcal{K}$, the deviation of the process entropy in presence of the stimulus from the entropy value in the absence of stimulus. This information gain is defined using the Kullback-Leibler divergence \cite{K-L} of the probability density $\overline{\Sigma}_n$  for events occurring in the presence of stimulus from the probability density $\Sigma_n$ for events occuring in the stationary condition without stimulus. This information measure is independent of the resolution $\Delta \tau$ and is given by
\begin{align}
\label{Ing0}
&{\mathcal K}(t,t_0)= \Sigma(t,t_0)\ln
\left(\frac{\Sigma(t,t_0)}{\overline{\Sigma}(t,t_0)}\right) + \\
&\nonumber  \sum_{n=1}^{\infty} \int_{t_0}^t
dt_1\int_{t_1}^t dt_{2} \cdots \int_{t_{n-1}}^t dt_n
\Sigma_n(t,t_n,\dots,t_0)\ln\left(\frac{\Sigma_{n}(t,t_n, \dots, t_0)}{\overline{\Sigma}_{n}(t,t_n,\dots,t_0)}\right).
\end{align}
Our goal is to evaluate, for the signaling module introduced in Sec. 2, the information gained about the input stimulus $V(t)$ from the knowledge of the output $\sigma(t)$.
\begin{figure}[h!]
\hspace{2cm}\includegraphics[height=5.9 cm,width=7.5 cm, angle=0]{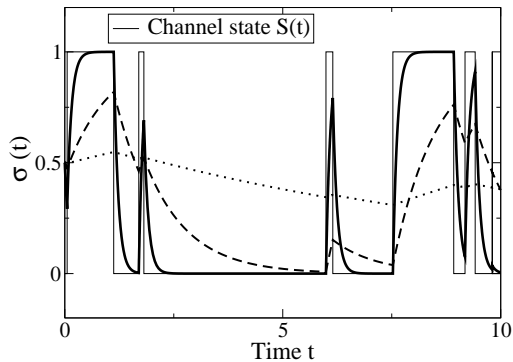}
\caption{Examples of time courses for the dimensionless calcium  concentration $\sigma(t)$ following from Eq. (\ref{dynamics}) for a given sequence $\vec{t}_n$ of channel state switches.  Thick, dashed and dotted lines correspond to $\lambda=10.0,1.0$ and $0.1$  respectively.}
\end{figure}
 In the module of Fig 1, given an initial concentration $\sigma_0$,  to each realisation, or trajectory, of $S(t)$
corresponds univocally one trajectory  of $\sigma(t)$ (see Fig. 2).
The trajectories of  $\sigma(t)$  can be therefore grouped in four sets determined
by the initial  and  final channel state. 
Let  
$\Sigma_{ij}(\vec{t}_n,t)$ be the probability of having a given realisation of $\sigma(t)$  in correspondence of  $n$ opening and closing  events of the channel occurring between $0$ and $t$ (indicated with $\vec{t}_n$ for brevity),
 with  $i,j=\pm$  indicating  the initial and final  channel state respectively ($+$ meaning open and $-$ closed).  
 We introduce then     the analogous probability for the unperturbed channel  dynamics $\Sigma_{ij}^0(\vec{t}_n,t)$,  i.e. corresponding to have neither stimulus nor feedback applied and we use   $\overline{\Sigma}_{ij}(\vec{t}_n,t)$ 
 for  the  dynamics including feedback  but not the stimulus.
For the unperturbed process  the waiting times distributions (WTDs) in each of the two states are Poissonian, i.e. $\psi_{\pm}^0(\tau)=\gamma_{\pm}e^{-\gamma_{\pm}\tau}$ and the corresponding survival probabilities $\Psi^0_{\pm}=\int_t^\infty \psi^0_{\pm}(\tau)d\tau=e^{-\gamma_{\pm} \tau}$. It follows that the  probability $\Sigma^0_{ij}(\vec{t}_{n},t)$ of  a time realisation of $\sigma(t)$   with 
a sequence  $\vec{t}_n$ of channel switches  beginning in state $i$ of the channel and ending  in  state $j$ is (in absence of stimulus and feedback):
\begin{align}
\nn 
\Sigma^0_{ij}(\vec{t}_{n},t)=P_{i}(\sigma_0)
\psi^0_{-i}(\tau_1)\psi^0_{i}(\tau_2)\cdots \psi^0_{-j}(\tau_n)\Psi^0_{j}(t-t_{n})
\end{align}
where  $\tau_k=t_k-t_{k-1}$ and $P_i( \sigma_0)$ is  the probability that at time $t=0$
the gate is in state $i$ and the initial concentration is $\sigma_0$. 

Eq. (\ref{Ing0})  applied to our module of Fig. 1 with the notation introduced above, becomes
\begin{align}
\label{Ing}
&{\mathcal K}[\sigma|V]= \sum_{i,j=\pm}\left[ \delta_{ij} \Sigma_{ii}(t)\ln
\left(\frac{\Sigma_{ii}(t)}{\overline{\Sigma}_{ii}(t)}\right) +\right. \\
&\nonumber \left. \sum_{n=1}^{\infty} \int_{t_0}^t
dt_1\int_{t_1}^t dt_{2} \cdots \int_{t_{n-1}}^t dt_n
\Sigma_{ij}(\vec{t}_n,t)\ln\left(\frac{\Sigma_{ij}(\vec{t}_n,t)}{\overline{\Sigma}_{ij}(\vec{t}_n,t)}\right)\right],
\end{align}
which, since there is a one-to-one relation between the output trajectories $\sigma(t)$ and the sequences of channel switches, 
 expresses the difference in uncertainty about the calcium fluctuations in the absence and presence of the input.  Note that Eq. (\ref{Ing}) does not include any averaging over inputs; rather, it is the information gain for a specific input realization $V(t)$. 
  \section{Evaluation of the information gain}
In this section, we carry out a perturbative calculation of the signal-induced information gain in the presence of feedback.
 For a switching process without feedback, the information gain was derived previously in \cite{Hanggi}. To evaluate Eq. (\ref{Ing}) in our case, the probabilities $\Sigma_{ij} (\vec{t}_n,t)$ and $\overline{\Sigma}_{ij} (\vec{t}_n,t)$ must include the feedback contributions to the desired order in $\eps_{\sigma}$.

Perturbation due to  either the stimulus or feedback leads 
to time-dependent rates and consequently  modified 
 survival probabilities $ \Psi_{\pm}(t,t_k)=\exp[{-\int_{t_k}^t \tilde{\gamma}_{\pm}dt'}]$ in each of the two channel states.
 The ensuing   processes are non-Poissonian, since the perturbed rates $\tilde{\gamma}_{\pm}$ depend on time, with  WTDs $\psi_{\pm}(t,t_k)=-d\Psi_{\pm}(t,t_k)/dt$ which  for small perturbation can be expanded as:
\begin{subequations}
\label{pwtd}
\begin{flalign}
&\psi_+(\tau,t_k)=\psi_+^0(\tau) \left[1+ \epsv \psi^V(\tau,t_k)+\epsv^2 \psi^{V^2}(\tau,t_k) \cdots \right]\\
&   \psi_-(\tau,\vec{t}_k)=\psi_-^0(\tau)\left[1+ \epss \psi^{\sigma}(  \tau,\vec{t}_k)+\epss^2 \psi^{\sigma^2}(\tau,\vec{t}_k)\cdots \right],
\end{flalign}
\end{subequations}
where $\tau=t-t_k$. The first order corrections are:
\begin{subequations}
\label{psiVs}
\begin{align}
&\psi^V(  \tau,t_k)=V(t_k+  \tau)-  \gamma_+ \int_{t_k}^{t_k+  \tau}dx V(x) \\
 & \psi^{\sigma}(  \tau,\vec{t}_k)=\sigma(t_k+  \tau|\vec{t}_k)-  \gamma_- \int_{t_k}^{t_k+  \tau}dx \sigma(x|\vec{t}_k) 
\end{align}
\end{subequations}
and the corresponding expressions for higher order corrections can be analogously derived. The appearance  of $\vec{t}_k$ in (\ref{psiVs}b)
remarks the dependence of this correction on all the sequence of events up to time $t_k$.  
Due to the form of Eqs. (\ref{psiVs})  and of the analogous higher order corrections, the WTDs in Eqs. (\ref{pwtd}) remain normalized at each order (the perturbative corrections integrate to zero  as can be checked by applying integration by parts on the first order corrections while using Eqs. (\ref{psiVs})).
As the integral of $\sigma(t)$ in Eq. (\ref{psiVs}b) can diverge  with $\tau$,   the validity of the small perturbation expansion for  $\psi_-$ has to be examined. The terms of the expansion of the information gain ensuing from the adoption of Eqs. (\ref{psiVs}) can be proved to be well behaved  when averaged over sequences $\vec{t}_n$  (see   Appendix, last paragraph).
Carrying out the  expansion to second order according to the Eqs. (\ref{pwtd})  and (\ref{psiVs}) leads to: 
\begin{align}
\label{fundterm}
 \Sigma_{ij}(\vec{t}_n,t)
&\nn \simeq \Sigma^0_{i j}+\epsv \Sigma^V_{ij}  + \epss \Sigma^{\sigma}_{ij}+\epsv^2
\left(\Sigma^{VV}_{ij}+\Sigma^{V^2}_{ij}\right)+\\
&+\epss \epsv \Sigma^{\sigma V}_{ij}+\eps_{\sigma}^2 \left(\Sigma^{\sigma
\sigma}_{ij}+\Sigma^{\sigma^2}_{ij}\right)
\end{align}
where to lighten notation we have dropped the dependence on $(\vec{t}_n,t)$  in the terms on the right-hand side. 
In Eq. (\ref{fundterm}) the first order  contributions to  the expansion for small $\epsv,\epss ($indicated  by superscripts $V$ and $\sigma$)
 are obtained by adding to   $\Sigma^0_{ij}(\vec{t}_n,t)$  the
corrections obtained drawing only one interval 
 in the sequence $\vec{t}_n$
 with a WTD  perturbed up to  the first order correction in  Eqs. (\ref{pwtd}) 
  while the remaining intervals are generated   with unperturbed  WTDs $\psi^0_{\pm}(\tau)$. 
 The  second order corrections to $ \Sigma_{ij}(\vec{t}_n,t)$  are obtained either by  drawing   two intervals in the sequence $\vec{t}_n$  with WTDs  corrected to first order  (superscripts $VV,\sigma V, \sigma \sigma$)   
 or by drawing only  one  interval in $\vec{t}_n$, 
but  with WTDs  corrected up to second  order (superscripts $V^2, \sigma^2$),  with the remaining intervals  drawn with  unperturbed WTDs $\psi^0_{\pm}(\tau)$.

Considering the probability $ \Sigma_{ij}(\vec{t}_n,t)\equiv Q_{ij}(\epss,\epsv)$  as
function of the small parameters $\epss, \epsv$, due to the structure of Eq. (\ref{tauent})
 the information gain  is a sum
of terms with the following functional form:
\BEQ 
\label{infgainbasic}
G_{ij}(\epss,\epsv)=Q_{ij}(\epss,\epsv) \ln \frac{Q_{ij}(\epss,\epsv)}{Q_{ij}(\epss,0)}.
\EEQ
A Taylor expansion of $G_{ij}(\epss,\epsv)$ to second order around $\epss=\epsv=0$ leads to:
\begin{align}
\label{taylor}
&G_{ij}(\epss,\epsv)=\epsv \frac{\partial Q_{ij}(\epss,\epsv)}{\partial \epsv}|_{\epss=\epsv=0}+\epss \epsv \left[ \frac{\partial^2 Q_{ij}(\epss,\epsv)}{\partial \epss \partial \epsv}+  \right. \\
& \nn \left.  \frac{ \frac{\partial Q_{ij}(\epss,\epsv)}{\partial\epss}\frac{\partial Q_{ij}(\epss,\epsv)}{\partial\epsv}}{Q_{ij}(\epss,\epsv)}\right]_{\epss=\epsv=0}+\frac{\epsv^2}{2}\left[\frac{\partial^2 Q_{ij}(\epss,\epsv)}{\partial \epsv^2}+\frac{ \left(\frac{\partial Q_{ij}(\epss,\epsv)}{\partial \epsv}\right)^2}{Q_{ij}(\epss,\epsv)}\right]_{\epss=\epsv=0}
\end{align}
with all missing second-order terms having coefficient equal to zero when evaluated at  $\epss=\epsv=0$ (for the full formal expansion see Appendix).
Replacing the expression of $Q_{ij}$ in terms of  $\Sigma_{ij}$ using Eq. ($\ref{fundterm}$),  one obtains for the second order expansion of  $G_{ij}$ the following terms:
\begin{align}
\label{expref}
\nn&
G_{ij}(\epss,\epsv)\simeq\epsv \Sigma_{ij}^V (\vec{t}_n,t)+\epss \epsv \Sigma^{\sigma V}_{ij}(\vec{t}_n,t)
+\\
&+ \epsv^2 \left[\Sigma^{V^2}_{ij}(\vec{t}_n,t)+\Sigma^{VV}_{ij}(\vec{t}_n,t)+\frac{\Sigma^V_{ij}(\vec{t}_n,t)^2}{2 \Sigma^0_{ij}(\vec{t}_n,t)}\right]
\end{align}
with the contribution of order $\epss^2$ being exactly zero.
As already mentioned, Eqs. (\ref{pwtd}) imply that the  corrections to the unperturbed
WTDs  integrate to exactly zero over time, so that the  WTDs remain normalized.
As a consequence $ \Sigma_{ij}^0 (\vec{t}_n,t)$, when
averaged over all possible sequences  $\vec{t}_n$ and initial and final channel state,  adds up to one, while all
the perturbative corrections of each order give a zero overall contribution.
 It follows therefore that  integrating and summing Eq. (\ref{expref})  
over all possible paths $\vec{t}_n$ and initial and final state $i,j$
the linear term $\Sigma^V_{ij}$, and the second-order terms $\Sigma^{\sigma V}_{ij}$, $\Sigma^{V^2}_{ij}$ and $\Sigma^{VV}_{ij}$ give zero contribution.

Let us then analyse the remaining $2^{nd}$-order term:  
$A^{(2)}_{ij}\equiv \frac{(\Sigma^V_{ij})^2}{2 \Sigma^0_{ij}}.$

Considering for example the case of initial and final closed state $i=j=-$, we obtain
\begin{align}
\label{amm1}
\nn A^{(2)}_{--}(\vec{t}_{n|k},t)&=A^{(2)}_{--}(t,t_n,\cdots t_k \cdots t_1,t_0)=\frac{\Sigma^V_{--}(\vec{t}_{n|k},t)^2}{2 \Sigma^0_{--}(\vec{t}_{n,k},t)}=
\\=&\frac{P_-(\sigma_0)}{2}\psi^0_+(\tau_1)\psi^0_-(\tau_2) 
\cdots
{\psi}^{V}(\tau_{k+1},t_k)^2\cdots\psi^0_-(\tau_{n}) \Psi^0_+(t-t_{n}) ,
\end{align}
where
$\Psi_{\pm}^0$ are
unperturbed survival probabilities
for the channel opening/closing events.
Further subscript $k$  in $\vec{t}_{n|k}$  indicates the perturbed time interval within the set of $n$ time intervals, 
with $k$ and $n$ in (\ref{amm1}) being even integers  for $i=j=-$.

For the distribution of initial conditions $P_{\pm} (\sigma_0 )$, it is natural to assume a stationary solution  for channel dynamics with feedback but no stimulus. 
   In this case the initial probabilities are $P_{\pm}(\sigma_0)=\gamma_{\pm}/(\gamma_+ +\gamma_-)
 \pm \Sigma_{\sigma}$, where a linear correction $\Sigma_{\sigma}\propto
\eps_{\sigma}$ is added to the stationary solution for the 
 unperturbed dynamics without feedback.
It follows that the correction due to feedback in $P_{\pm}(\sigma_0)$ produces in Eq. (13) a contribution of order $ \epsv^2 \epss $, and can therefore  be neglected to second order.  Moreover, a direct calculation shows that this 
correction is
$\propto \epsv^2 \eps_{\sigma}\gamma_+ e^{-t(\gamma_++\gamma_-)} V^2(t)$ and therefore vanishes on the channel state switching time-scale.
It can consequently be neglected also to third order when evaluating information gain over long time intervals.
Averaging over all possible paths,  the second order contribution $A_{ij}$ to the information gain is:
\begin{align}
\label{hangres}
\mathcal{K}^{(2)}(t)&=\sum_{i j,n k}\int d \vec{t}_{n|k}
A^{(2)}_{ij}(\vec{t}_{n|k},t)=\\
&\nn=\epsv^2  \int d\omega d\omgp \hat{V}(\omg)\hat{V}(\omgp)\frac{\imath \gamma_+
\gamma_-[1-e^{\imath (\omg+\omgp)t}]}{2(\omg+\omgp)(\gamma_-+\gamma_+)}
\end{align}
where the sum over $n$ runs from $1$ to $+\infty$ ($k<n$) and
for convenience of calculation we introduced the Fourier transform $\hat{V}(\omega)$ 
\BEQ
\hat{V}(\omega)=\int_{-\infty}^{\infty} dt e^{-\imath \omega t} V(t)
\EEQ
  of the stimulus $V(t)$, adopting the character $\imath$ for the imaginary unit.
Taking the time derivative 
and integrating over frequencies gives the rate of information gain:
\BEQ
\label{finallowest}
\frac{d \mathcal{K}}{dt}=\frac{\epsv^2}{2} V^2(t) \frac{\gamma_+\gamma_-}{\gamma_+ + \gamma_-}
\EEQ
which 
coincides with  the result  obtained in   \cite{Hanggi} for a similar signaling module without feedback.

In order to see the contribution of the feedback one then has to expand
 Eq. (\ref{expref}) to third  order. This leads to the additional terms shown in Eq. (\ref{3order}). After summation, integration (for details see Appendix) we obtain
 \begin{align}
 &  \label{hangres2Z}\mathcal{K}^{(3)}(t)=\frac{1}{2}\sum_{ij,nkl}\int d \vec{t}_{n|kl}
A^{(3)}_{ij}(\vec{t}_{n|kl},t)= -\frac{\imath
\epsv^2 \epss}{2} \frac{\gamma^2_+ \gamma_-}{(\gamma_+ +\gamma_- )^2}\\
 \nn& \int d\omg d\omgp \frac{\hat{V}(\omg)\hat{V}(\omgp)}{\omg+\omgp} \frac{\gamma_+ +\lambda}{\gamma_- +\gamma_+ +\lambda} e^{\imath(\omg +\omgp)t}.
 \end{align}
We differentiate (\ref{hangres2Z}) with  respect to time, integrate over frequencies and  add up to (\ref{finallowest})  so as to obtain the information gain rate including the feedback contribution:
\BEQ
\label{finalstat}
 \frac{d \mathcal{K}}{dt}=\frac{\epsv^2}{2}V^2(t)\frac{\gamma_+ \gamma_-}{\gamma_+
+\gamma_- }\left[1+\epss\frac{\gamma_+ }{\gamma_+
+\gamma_- }\frac{\gamma_+ + \lambda}{\gamma_+ +\gamma_-+\lambda}\right] ,
\EEQ
where terms vanishing on time scale $t\gg \mbox{max}(\lambda^{-1},\gamma^{-1}_{\pm})$, which include those carrying the dependence on the initial concentration $\sigma_0$,
 have been neglected.
Our calculation was carried out for the case of the input and the feedback acting on the opening and closing rate, respectively. Our approach can obviously be extended to the case of interchanged action of feedback and input on the rates, which would lead to the same result as in Eq. (\ref{finalstat}) with the simple interchange $\gamma_+ \leftrightarrow \gamma_-$. In the case of both feedback and input affecting the same rate, the leading feedback correction would again be of order $\epss \epsv^2$, but the coefficient would be different. The terms of order e.g. $\epss \epsv$  in the derivation can in this case emerge from perturbing to second order the same switching event, while in the case we analyzed they can only emerge from perturbing two switching events to first order. Derivation of the correction for this case, as well as of higher-order corrections, is beyond the scope of this paper.

\section{Discussion}
The information gain rate Eq. (\ref{finalstat}), when integrated over a time interval $[0,T]$ with $T \gg $max$(\lambda^{-1}, \gamma_+^{-1})$, gives the total information obtained about the stimulus by observing the output. It is seen that (to first order in feedback strength and second order in stimulus strength) the information gain depends only on the total power $\int_0^T dt V^2(t)$  of the stimulus, rather than on the spectral distribution of this power.
This implies that the information gain rate cannot be optimized by matching the temporal structure of the stimulus with the time scale of the feedback dynamics.

The factor $ \frac{\gamma_+ \gamma_-}{\gamma_+  + \gamma_- }$  in Eq. (\ref{finalstat}) expresses the rate of 'double flip events' (i.e., openings and successive closings of the channel) in the absence of stimulus and feedback. In the case $\eps_\sigma=0$, the information gain rate is therefore simply proportional to the unperturbed rate of double flip events. \footnote{The true double flip rate in the presence of stimulus differs from $\frac{\gamma_+ \gamma_-}{\gamma_++\gamma_- }$ by a correction that is first order in $\eps_V$. In Eq. (\ref{finalstat}), such a correction would result in a higher-order term $o(\eps_{\sigma}^3)$, which is beyond the order in which we carried out the expansion.}To examine if an analogous relation holds generally, we evaluate the mean rate of double flip events in absence of stimulus but in presence of feedback. The mean closing rate is given by $\gamma_- (1+\eps_\sigma \overline{\sigma}_+$), where $\overline{\sigma}_+$ is the average concentration when the channel is in the open state. To obtain a result valid to first order in $\eps_\sigma$, it is sufficient to express $\overline{\sigma}_+$  to $0^{th}$ order; this conditional average was computed in Ref.  \cite{boro} and equals 
\BEQ
 \overline{\sigma}_{+}=\frac{\gamma_-+\lambda}{\gamma_-+\gamma_+ +\lambda}.
 \EEQ
 The mean rate of double flip events in the presence of feedback is then given by
\begin{align} 
R&=\gamma_+ \gamma_- (1+\eps_{\sigma} \overline{\sigma}_{+})/(\gamma_++\gamma_-(1+\eps_{\sigma} \overline{\sigma}_{+})) \\
\nonumber &= \frac{\gamma_- \gamma_+}{\gamma_-+\gamma_+ } \left[1+\eps_{\sigma} \frac{ \gamma_+}{\gamma_-+\gamma_+ }\frac{\gamma_++\lambda}{\gamma_-+\gamma_+ +\lambda}+o(\eps_{\sigma}^2)\right]
\end{align}
and Eq. (\ref{finalstat}) becomes
\BEQ
\label{finaleq}
\frac{d K}{dt}=\frac{\eps_V^2}{2} V^2 (t)R.
\EEQ
The information gained per double flip event is therefore given only by the power of the input and does not depend on the feedback strength $ \eps_{\sigma}$  or on the kinetic parameters $\gamma_+ $,$\gamma_-$,$\lambda$.

Before relating this result to findings from the previous literature, we first point out that the definition of �feedback� in communication / information theory is more restrictive than the definition used in the literature on biochemical signaling. In the biological literature, feedback arises when a product of a signal transduction step influences the upstream components in the pathway. This fits with the general definition of feedback in the early cybernetics literature: "When (this) circularity of action exists between the parts of a dynamic system, feedback may be said to be present" \cite{ashby}. In communication theory, however, feedback is typically required to act in such a way that it effectively modifies the input of the system.  The signaling module analyzed in this work contains in fact a feedback loop in the former sense; functionally, such type of feedback permits to e.g. achieve sensory adaptation to repeated stimuli \cite{reidl} and can improve the temporal resolution of signaling \cite{boro}. In the latter sense (viewed as a communication  channel), however, our module cannot be said to have feedback, as the input $V(t)$ is not combined with the output. Rather, the back-coupling implements an autoregulatory loop with the communication channel. The effect of the autoregulatory loop is to  give memory to the channel state-switching dynamics: the closing rate depends (through the instantaneous calcium concentration) on the full history of previous channel opening and closings. Our signaling module can therefore be viewed as a non-Markovian point process channel without feedback.
 
Following this clarification of terminology, to put  the results of Eqs.(\ref{finalstat}) and (\ref{finaleq}) into perspective,  we recall a known result from information theory
It was proved in \cite{davis, kabanov} (see also \cite{johnson}), that for any   Markov point process the channel capacity per event cannot exceed the capacity of the Poisson process. I.e., the capacity per event is not improved by memory. 
 While we have calculated the information gain rate for one realization of the stimulus (rather than the channel capacity), our finding may be viewed as somewhat analogous. In our case, the channel is non-Markovian, but as the memory decays exponentially in time (akin to a Hawkes process), the system is 'near-Markov'. It is possible, however, that a different result (i.e., information gain affected by memory) would be obtained if we carried out the perturbation expansion to higher orders in $\eps_{\sigma}$. For discrete-time (rather than point-process) channels, the influence of memory on channel capacity was recently analyzed in \cite{kostal,kostalb}.
 

In \cite{tenwoldepre}, an analysis of several prototypical signaling pathways with feedback was carried out under the additive Gaussian noise approximation. Under this restriction (see also Sec. 1), the authors were able to compute the mutual information rate for modules in which nonlinear feedback affected the activation of an upstream component. They concluded that when the feedback was mediated by the final output of the pathway, no improvement of the information transmission was obtained. This  is reminiscent of our main finding. In \cite{tenwoldepre}, an enhancement of information transmission was obtained only when the feedback was mediated by an intermediate product in the pathway, and not the final output. The module we analyzed in this paper (Fig. 1) falls outside of this class.

In conclusion, we presented a novel perturbative approach that permits to analytically evaluate information transfer through non-Markovian point process channels. We applied this approach to a prototypical signaling module with biochemical feedback and showed that  to first order in feedback strength  the information gain rate is increased by negative feedback (and decresead by positive feedback). However, this change in information gain rate is fully accounted for by the feedback-induced change in the rate of signaling events (channel opening/closings). To first order in feedback strength, the information gain per signaling event is not affected by feedback.

\section{Acknowledgments}
We thank Lubomir Kostal for insightful comments on the manuscript. M.Z. acknowledges institutional support RVO:67985823 and grant support P304/12/G069 (The Czech Science Foundation).





\section*{Appendix: Expansion of the Information gain }
The Taylor expansion of Eq. (\ref{infgainbasic})  to second order  in $\epss, \epsv$ gives

\begin{align}
\label{taylor}
\tag{A1}
&G_{ij}(\epss,\epsv)=\epss\left[ \frac{\partial Q_{ij}(\epss,\epsv)}{\partial \epss}
- \frac{Q_{ij}(\epss,\epsv)}{Q_{ij}(\epss,0)}\frac{\partial Q_{ij}(\epss,0)}{\partial \epss}\right]_{\epss,\epsv=0}\\
\nn&+\epsv \frac{\partial Q_{ij}(\epss,\epsv)}{\partial \epsv}|_{\epss,\epsv=0}+\frac{\epss^2}{2}\left[\frac{\partial^2 Q_{ij}(\epss,\epsv)}{\partial \epss^2} +\left(\frac{\partial Q_{ij}(\epss,\epsv)}{\partial \epss}\right)^2 \frac{1}{ Q_{ij}(\epss,\epsv)}\right. \\
\nn&\left.+\frac{ Q_{ij}(\epss,\epsv)}{Q^2_{ij}(\epss,0)}\left(\frac{\partial Q_{ij}(\epss,0)}{\partial \epss}\right)^2  -\frac{2}{ Q_{ij}(\epss,0)}\frac{\partial Q_{ij}(\epss,\epsv)}{\partial \epss}\frac{\partial Q_{ij}(\epss,0)}{\partial \epss}\right.\\
\nn &\left. - \frac{\partial^2 Q_{ij}(\epss,0)}{\partial \epss^2}\frac{Q_{ij}(\epss,\epsv)}{Q_{ij}(\epss,0)} \right]_{\epss,\epsv=0}+\frac{\epsv^2}{2}\left[\frac{\partial^2 Q_{ij}(\epss,\epsv)}{\partial \epsv^2}+\frac{1}{Q_{ij}(\epss,\epsv)} \left(\frac{\partial Q_{ij}(\epss,\epsv)}{\partial \epsv}\right)^2\right]_{\epss=\epsv=0}\\
\nn &+\epss\epsv\left[\frac{\partial^2 Q_{ij}(\epss,\epsv)}{\partial \epss \partial \epsv}
+\frac{1}{Q_{ij}(\epss,\epsv)} \frac{\partial Q_{ij}(\epss,\epsv)}{\partial\epss}\frac{\partial Q_{ij}(\epss,\epsv)}{\partial\epsv}\right]_{\epss=\epsv=0},
\end{align}
from which it is readily seen that the coefficient of the term $\propto \epss$ and $\propto \epss^2$ vanish when evaluated at $\epss=\epsv=0$.

Expanding Eq. (\ref{infgainbasic}) to third order, following the same procedure as in Eq. (\ref{taylor}) and converting back  $Q_{ij}$ to $\Sigma_{ij}$ 
leads to three additional terms:
\begin{align}
\label{3order}
\tag{A2}
&\nn \frac{ \epss^2 \epsv }{3}\left[\Sigma_{ij}^{\sigma \sigma V}   +\Sigma_{ij}^{\sigma^2 V}  \right]+\frac{\epsv^2 \eps_{\sigma}}{6} \left[4
\frac{\Sigma_{ij}^{V}  \Sigma_{ij}^{\sigma
V}   }{\Sigma^0_{ij}  } -
\frac{ (\Sigma_{ij}^{V} )^2
\Sigma_{ij}^{\sigma}  }{(\Sigma^0_{ij} )^2}
\right]\\
\nn &+\frac{\eps_V^3}{6} \left[-\frac{(\Sigma_{ij}^{V} )^3}{
(\Sigma^0_{ij} )^2} +6\Sigma_{ij}^{V}  \frac{\Sigma_{ij}^{VV}  +\Sigma_{ij}^{V^2}   }{\Sigma^0_{ij}  }\right].
\end{align}
where we have used the next order expansion of (\ref{fundterm}) i.e.
\begin{align}
\label{fundterm3}
\tag{A3}
 &\Sigma_{ij}(\vec{t}_n,t)
 \simeq \Sigma^0_{i j}+\epsv \Sigma^V_{ij}  + \epss \Sigma^{\sigma}_{ij}+\epsv^2
\left(\Sigma^{VV}_{ij}+\Sigma^{V^2}_{ij}\right)+\epss \epsv \Sigma^{\sigma V}_{ij}\\
&\nn+\epss^2 \left(\Sigma^{\sigma
\sigma}_{ij}+\Sigma^{\sigma^2}_{ij}\right)+\epss^3 \left(\Sigma^{\sigma
\sigma \sigma}_{ij}+\Sigma^{\sigma
\sigma^2}_{ij}+ \Sigma^{\sigma^3}_{ij}\right)+\epss^2 \epsv \left(\Sigma^{\sigma
\sigma V}_{ij}+\Sigma^{\sigma^2 V}_{ij}\right)\\
& \nn\epss \epsv^2 \left(\Sigma^{\sigma
VV}_{ij}+\Sigma^{\sigma V^2}_{ij}\right)+\epsv^3 \left(\Sigma^{V
VV}_{ij}+\Sigma \Sigma^{V
V^2}_{ij}+ \Sigma^{V^3}_{ij}\right)
\end{align}
The third term in Eq. (\ref{3order}) depends only on the stimulus while the first, proportional to $ \epss^2 \epsv$, is exactly 
the same coefficient that would appear as the third order in the expansion in Eq.  (\ref{fundterm}) and therefore
gives zero contribution after integration over $\vec{t}_n $ and summation,  since perturbative corrections do not affect the normalization of the unperturbed part.
Therefore  the contribution with the feedback is given only by the term  proportional to $ \epsv^2  \epss$ in Eq. (\ref{3order}),
whose  first element  can be written as:
\begin{align}\label{rearrange}\tag{A4}
 &\frac{\Sigma_{ij}^{V}\Sigma_{ij}^{\sigma V} }{\Sigma^0_{ij}}=\Sigma_{ij}^{\sigma V 
V}+\Sigma_{ij}^{\sigma}
\frac{\left(\Sigma_{ij}^{V}\right)^2 }{\Sigma^0_{ij}} =\left[\Sigma_{ij}^{\sigma V 
V}+\Sigma_{ij}^{\sigma
V^2}\right]+\Sigma_{ij}^{\sigma}
\frac{\left(\Sigma_{ij}^{V}\right)^2 }{\Sigma^0_{ij}} -\Sigma_{ij}^{\sigma V^2}
\end{align}
The term between brackets  has been so rearranged in order to show that after summation and integration, it 
gives zero contribution (since it  corresponds to the
term of order $\epsv^2 \epss$  in the
the expansion of Eq. (\ref{fundterm}), i.e. first term in last row in Eq. (\ref{fundterm3})).
The last term in (\ref{rearrange}) gives a contribution that vanishes on the time scale of the channel opening/closing dynamics.
The remaining term  combines with the second element in the feedback contribution in (\ref{3order})
taking an overall $1/2$ coefficient. The expression for this term, evaluated between initial and final closed states, is:
\begin{align}
\label{feedbackex}\tag{A5}
&A^{(3)}_{--}(\vec{t}_{n|kl},t)=\Sigma_{--}^{\sigma} \left(\frac{\Sigma_{--}^{V}}
{\Sigma^0_{--}}\right)^2=\\
\nonumber &=P_-(\sigma_0) \psi^0_+(\tau_1)\psi^0_-(\tau_2)
 \cdots {\psi}^{V}(\tau_{k+1},t_k)^2\cdots \psi^{\sigma}(\tau_{l+1}|\vec{t}_{l})
\cdots\psi^0_-(\tau_{n}) \Psi^0_+(t-t_{n})
\end{align}
where further subscripts  $k$ and $l$ in $\vec{t}_{n|k,l}$ indicate   that  the closing and the opening events affected by the stimulus and the feedback are respectively   the $k^{th}$  and the $l^{th}$ of the $n$ event  occurring between $0$ and $t$.   $\psi^{\sigma}(\tau_{l+1}|\vec{t}_{l})$ 
is the first order term given by Eq. (\ref{psiVs}) and  carries the dependence on the 
 history of the process prior to $t_l$ (i.e. the all sequence $\vec{t}_l$ of  events up to $t_l$), 
but, due to the exponential form of the WTDs,
 does not affect the convolution structure in (\ref{feedbackex}).


Averaging over all possible paths gives:
\begin{align}
\label{hangres2}\tag{A6}
\nn &  \mathcal{K}(t)=\frac{1}{2}\sum_{ij,nkl}\int d \vec{t}_{n|kl}
A^{(3)}_{ij}(\vec{t}_{n|kl},t)= -\frac{\imath
\epsv^2 \epss}{2} \frac{\gamma^2_+ \gamma_-}{(\gamma_+ +\gamma_- )^2}\\
 \nn& \int d\omg d\omgp \frac{\hat{V}(\omg)\hat{V}(\omgp)}{\omg+\omgp} \frac{\gamma_+ +\lambda}{\gamma_- +\gamma_+ +\lambda} e^{\imath(\omg +\omgp)t},
 \end{align}
plus terms that either  do not depend on $t$ or vanish  for time $t\ll $max$(\lambda^{-1}, \gamma_{+}^{-1})$ exceeding the channel state switching time scales.

Finally a  remark on the validity of the perturbative approach we introduced.
This approach is based on a direct perturbation of the switching rates and therefore of the WTDs, due to both input and feedback.  The only delicate  point  in following this procedure is the integral of $\sigma(x|\vec{t})$ in Eq. (\ref{psiVs}b), which can diverge  with $\tau$. It can be verified, however that this correction is well-behaved after summation over all stochastic paths and integration.
In fact  we  can replace  the contribution of this integral to the first order correction in Eq. (\ref{psiVs}b) with  the term $-\gamma_-\psi_-^0(  \tau)  \tau$,  which is the upper limit to the correction due to the integral  between $0$ and $\tau$ according to our perturbative prescription. The Laplace transform of this correction is
\BEQ
\label{A6}
\tag{A6}
\mathcal{ L} [-\gamma_{-}^2t e^{-\gamma_- t}]=-\int_0^{\infty}dt\gamma_{-}^2t e^{-\gamma_-t}e^{-s t } =  \left(\frac{\gamma_-}{\gamma_-+s}\right)^2=\hat{\psi}^0_-(s)^2
\EEQ
where 
\BEQ
\label{A7}
\tag{A7}
\hat{\psi}^0_-(s)=\mathcal{ L} [\gamma_{-} e^{-\gamma_- t}]=\int_0^{\infty}dt\gamma_{-} e^{-\gamma_-t}e^{-s t } =\frac{\gamma_-}{\gamma_-+s}.
\EEQ
We can now replace Eq. (\ref{A6})  in the transition amplitude.  Due to the convolution structure of $\Sigma_{ij}(\tau, t)$ in terms of WTDs, this replacement adds just a  multiplicative factor in Laplace space. The final result is (for $i=j=-$)
\BEQ
\label{converge}
\tag{A8}
\Sigma_{--}(s)= \Sigma^0_{--}(s)(1+\epss \hat{\psi}^0_-(s)),
\EEQ
where we have included only the contribution of the integral and not  the first term in Eq. (\ref{psiVs}b) since the latter is trivially well-behaved.
Due to Tauberian theorem, one can deduce the time asymptotic limit by analyising the $s\to 0$ behavior in Eq. (\ref{converge}). Since $\hat{\psi}_-(s)\to 1$ for $s \to 0$, such correction obtained after averaging over trajectories is  always limited, non-divergent and therefore well-behaved for $t \to \infty$.

\section*{References}


\end{document}